# Dynamic clay microstructures emerge via ion complexation waves

Michael L. Whittaker,[1,2,*] David Ren,[3] Colin Ophus,[4] Yugang Zhang,[5] Benjamin Gilbert,[1,2] Laura Waller,[3] Jillian F. Banfield[1,2]

[1] *Energy Geosciences Division, Lawrence Berkeley National Laboratory, USA 94720.*
[2] *Department of Earth and Planetary Science, University of California, USA 94720.*
[3] *Department of Electrical Engineering and Computer Sciences, University of California, USA 94720.*
[4] *National Center for Electron Microscopy, Molecular Foundry, Lawrence Berkeley National Laboratory, USA 94720.*
[5] *Center for Functional Nanomaterials, Brookhaven National Laboratory, Upton, New York, USA 11973.*

(Dated December 15, 2020)

Clays control carbon, water and nutrient transport in the lithosphere[1-4], promote cloud formation[5] and lubricate fault slip[6-8] through interactions among hydrated mineral interfaces[9]. Clay mineral properties are difficult to model because their structures are disordered, curved[10] and dynamic[11]. Consequently, interactions at the clay mineral-aqueous interface have been approximated using electric double layer models based on single crystals of mica[12,13] and atomistic simulations[14]. We discover that waves of complexation dipoles at dynamically curving interfaces create an emergent long-range force that drives exfoliation and restacking over time- and length-scales that are not captured in existing models. Curvature delocalizes electrostatic interactions in ways that fundamentally differ from planar surfaces, altering the ratio of ions bound to the convex and concave sides of a layer. Multiple-scattering reconstruction of low-dose energy-filtered cryo electron tomography enabled direct imaging of ion complexes and electrolyte distributions at hydrated and curved mineral interfaces with ångstrom resolution over micron length scales. Layers exfoliate and restack abruptly and repeatedly over timescales that depend strongly on the counterion identity, demonstrating that the strong coupling between elastic, electrostatic and hydration forces in clays promote collective reorganization previously thought to be a feature only of active matter[15].

Keywords: clay, cryoET, electric double layer, curvature, complexation

Quantitative models of solution structure at charged interfaces are of nearly ubiquitous interest across the physical sciences[16] and are a particularly acute need for clays, whose behavior is governed almost entirely by interactions between hydrated interfaces that are sensitive to subtle variations in structure[17]. Measurements of the potential drop through the electrochemical double layer[18] (EDL), the change in orientation and density of water molecules[19-21], and ordering of water and counterions[22-26] have revealed departures from contemporary EDL models. However, direct images of electrolyte distributions where liquids and solids meet, the strongest constraint on theories of hydrated interfaces, have remained elusive.

In the absence of direct observations, many structural models have been proposed to explain the diverse and time-varying behavior of layered minerals observed over a wide range of conditions[27-30]. Colloidal interactions between hydrated clay layers are often described by Derjaguin and Landau, Verwey and Overbeek (DLVO) theory, or by atomistic simulations of idealized layer configurations. However, neither DLVO theory nor simulations generalize to layer ensembles typically encountered in natural environments[10,31], which are comprised primarily of defective stacks of curved layers that evolve over time[32-34]. Curvature is key to understanding the solution structure at clay interfaces, which we show here leads to emergent interactions whose consequences manifest over a range of spatial and temporal scales.

CryoET of lithium-montmorillonite (Li-Mt) suspensions (Fig. 1) with a mineral volume fraction of 2% and electrolyte concentrations of $C = 0.1$ M and 0.75 M consist of exfoliated layers approximately one nanometer thick that bend and stack into loose aggregates with a broad range of interlayer spacings. We used small defocus values relative to the dimensions of the sample in order to achieve high resolution, which required the implementation of a reconstruction algorithm that accounts for multiple electron scattering[35] from layers in edge-on orientations while preserving low-contrast information from face-on layers and aqueous electrolyte. This enabled 3.64 Å real-space isotropic resolution over a 1.02 μm x 0.79 μm x 0.36 μm field of view. Rich datasets, with structural information over four orders of spatial magnitude, provide an unprecedented view of the suspension microstructure.

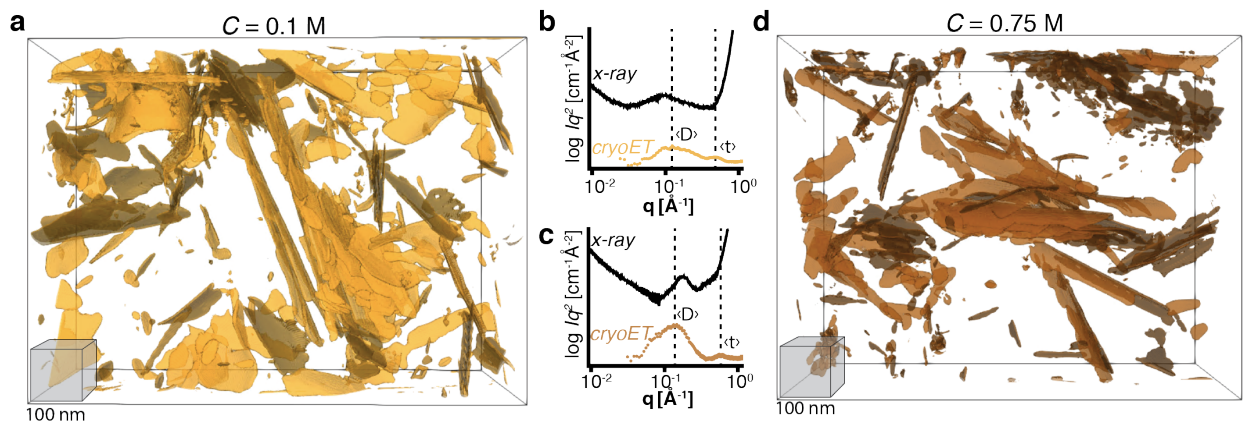

**Fig 1 | Li-Mt suspensions viewed with cryoET.** Isosurfaces of clay layers in 0.1 M (a) and 0.75 M (d) lithium chloride. (b, c) Comparison of structure factors from X-ray scattering and cryoET in in 0.1 M (b) and 0.75 M (c).

The large field of view in cryoET enables direct comparison to the structure factor determined by *in situ* X-ray scattering, providing a direct link between static nanoscale and dynamic mesoscale properties (Fig. 1b, c). A broad local maximum corresponding to the average interlayer spacing, $\langle D \rangle$, differs between the two techniques, indicating that cryoET subsamples of the larger clay suspensions contain a distinct distribution of interlayer spacings. Therefore, the most common structural parameter used to describe clay suspensions is not a high-fidelity representation of the true microstructure. As we describe in detail below, displacement of the mesoscale distribution in cryoET from the ensemble distribution observed in X-ray scattering is an indication that layer interactions are highly correlated over distances commensurate with the size of the reconstruction box and that the layers dynamically reorganize on length scales much larger than the average layer diameter.

CryoET reveals ion complexation distributions that are hidden in X-ray profiles by the high scattering intensity of water. A peak in the cryoET structure factor between scattering vectors $q = 0.50$-$0.57$ Å$^{-1}$ (Fig. 1b, c) corresponds to an average layer thickness, $\langle t \rangle$, that includes the aluminosilicate layer and hydrated lithium counterions but not bulk electrolyte. We determined that the thickness of an exfoliated layer is 12.6 Å in 0.1 M lithium chloride and 11.0 Å in 0.75 M lithium chloride. Lithium, and cations more generally, form complexes with negatively-charged Mt layers either in outer-sphere (fully hydrated) or inner-sphere (partially dehydrated) configurations[12]. Thicker layers at low electrolyte concentration are thus indicative of a greater fraction of fully hydrated lithium ions that reside further from the layer midplane, while conversely, thinner layers at elevated electrolyte concentration are direct evidence of a higher fraction of lithium ions that make inner-sphere complexes with the mineral interface. This is in accordance with the expectation of EDL models, which predict that higher electrolyte concentration drives complexation equilibria towards partial dehydration (Supporting Information).

Despite conceptual agreement between reciprocal-space cryoET structure factors and classical EDL models, real-space ion density profiles reveal a prominent role for curvature in modulating interfacial ion distributions. All layers are curved (shown in Fig. 2a-c for 0.1 M lithium chloride electrolyte), despite having chemical compositions that are nominally symmetrical about the layer midplane. Curved layers have approximately cylindrical symmetry, bending primarily along a single principal axis to generate convex and concave sides. We find that mean layer curvature, $H$, varies both within and among layers and is associated with asymmetric ion densities on opposing sides of a layer (Fig. 2d-f).

Relative electrolyte densities are quantified from the reconstructed absorbance magnitude. Between $5.3 \times 10^4$ and $1.3 \times 10^5$ absorbance profiles taken normal to layer at each midplane voxel revealed features beyond the resolution achievable directly from three dimensional tomographic images (Fig. 2). Key aspects of the solution structure expected for charged surfaces in an electrolyte solution are reproduced. For example, a region of low absorbance extends 5.0 nanometers from each interface when layers are nearly flat (Fig. 2b), arising from the depletion of chloride ions that are repelled from the negatively charged mineral. Thus, absorbance profiles moving from the layer midplane into the bulk solution can be attributed to dominant contributions from mineral, lithium, and chlorine, respectively.

Surprisingly, changing layer topology leads to significant ion density perturbations. The depletion region decreases in thickness and ions are concentrated in excess of their bulk solution densities on the concave side at distances up to 15 nm from the mineral interface (Fig. 2e). Thus, concavity sequesters electrolyte over appreciable distances. The key difference compared to planar interfaces is that electrostatic forces experienced by the electrolyte are delocalized because they depend on the integrated potential across the entire curved interface[36], which changes as complexation equilibria shift, and not simply a one-dimensional potential normal to an interface with constant charge, as typically envisioned in EDL models.

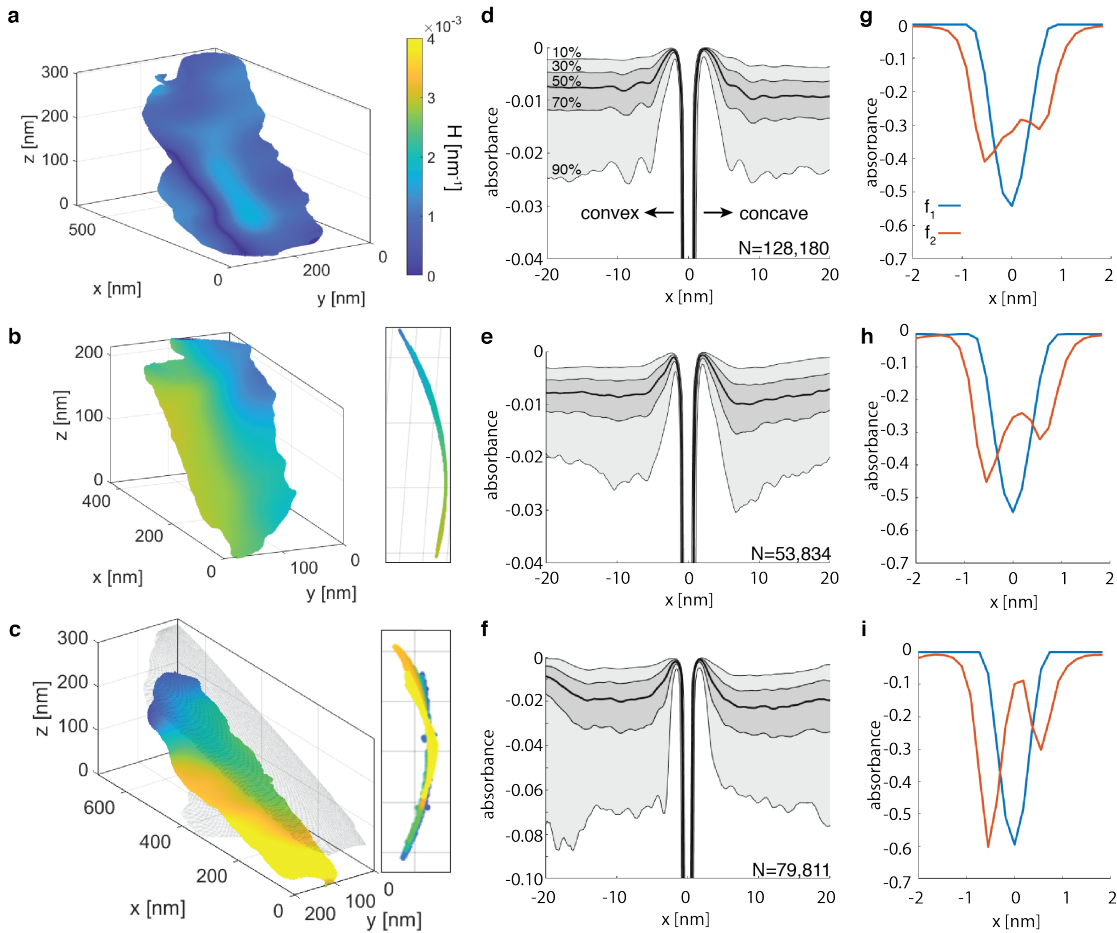

**Fig 2 | Effect of layer curvature on ion density profiles in 0.1 M lithium chloride.** (a) Layer with low curvature. (b) Layer with intermediate curvature. (c) Highly curved layer in an osmotic hydrate stack. (d) Generally symmetric absorbance profile within 20 nm of convex and concave sides of low-curvature layer. (e) Asymmetric absorbance profile on either side of curved layer. (f) Asymmetric, high magnitude absorbance profiles between stacked layers (note difference in scale). (g) NNMF of all absorbance profiles in (d), showing first two factors, $f_1$ and $f_2$. (h) Increasingly asymmetric $f_2$ in NNMF profile of curved layer in (b). (g). (i) Highly asymmetric NNMF $f_2$ for curved stacked layer.

We find that osmotic hydrates are stacks of layers with coaligned curvature axes that have higher electrolyte densities between them than exfoliated layers (Supporting Figure 1). In 0.1 M lithium chloride the average interlayer spacing is 20 nm and the median absorbance is over twice as high as a nearly planar layer (Fig 2c, f). As a result of the increased electrolyte concentration, osmotic hydrates have greater curvature and the interfacial complexation in osmotic hydrate layers is highly asymmetric, with the inner-sphere ion density on the convex side of the layer far higher than for isolated layers (Fig. 2i). This indicates that both curvature and interactions with neighboring layers influence ion complexation, which we explore in further detail at higher resolution.

Lithium ion complexation to the mineral interface was quantified by clustering real-space absorbance profiles. Using non-negative matrix factorization (NNMF) of all profiles for a given layer (Fig. 2c, f, i), we find that the first factor with 12.6 Å width to arise from the mineral layer and the associated outer-sphere complexation of hydrated lithium, confirming the reciprocal-space interpretation of the cryoET structure factor (Fig. 1c). Binding of partially dehydrated lithium, forming an inner-sphere complex with the layer, is captured by the second NNMF factor with a peak-to-peak width of 11.0 Å. This factor has greater absorbance on the convex side relative to the concave side and increases with increasing curvature (Fig. 2c, f, i), a clear demonstration that both inner- and outer-sphere complexation states coexist and that their relative proportions are dependent on layer curvature.

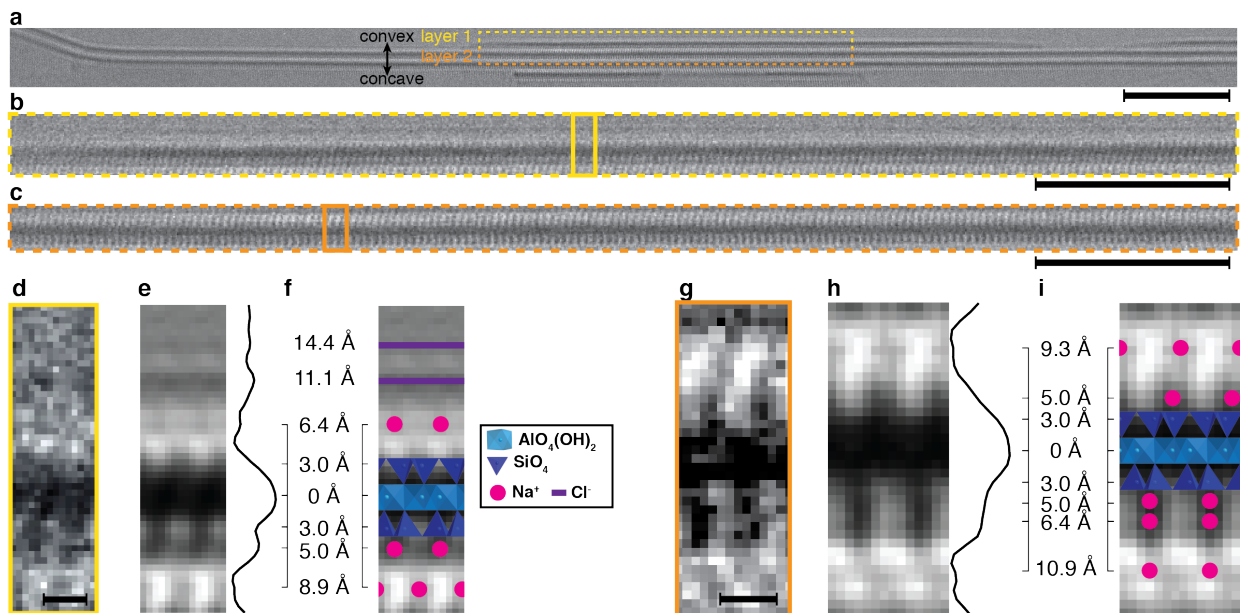

**Fig 3 | CryoEM structure of Na-Mt and aqueous interlayer solution.** (a) Na-Mt tactoid with four layers, each separated by 19 Å (scale bar represents 20 nm). (b) Layer 1 enlarged. (c) Layer 2 enlarged. (d) Two unit cell wide region from layer 1 showing the interface with the aqueous solution (scale bar represents 5 Å). (e) Average unit cell, duplicated for clarity. Normal intensity profile obtained from the average of the lateral intensities. (f) Structural model of external Mt layer overlain on average unit cells. (g) Two unit cell wide region from layer 2. (h) Average unit cell, duplicated for clarity. (i) Structural model of internal Mt layer.

Low-dose tomographic reconstructions had insufficient contrast to visualize hydrated lithium directly, but low-dose cryo electron microscopy (cryoEM) of sodium montmorillonite (Na-Mt) revealed ion distributions at the clay mineral interface with atomic resolution (Fig 3). In 1 M sodium chloride, curved clay layers stack into crystalline hydrates in which anions are excluded from the interlayer space[37] and the chloride ion depletion region extends over a Debye length of 3.1 Å at the external surface, allowing us to directly interpret the image contrast arising from interfacial complexation of sodium.

We focus on two layers within a stack, labeled (1) (Fig. 3b, d), and (2) (Fig. 3c, g), that highlight external and internal interfaces with both convex and concave sides. Tetrahedral cations are consistently found 3.0 Å along the normal direction from the octahedral cations at the layer midplane, as expected for Mt, serving as a reference for hydrated ion configurations (Fig. 3f, i). In layer (1), the average unit cell has primarily one-dimensional structure normal to the midplane with little lateral ordering on the externally facing side (Fig. 3e), while sodium ions are ordered within the interlayer space on the opposing side (Fig. 3h). In the interlayer, sodium formed primarily inner-sphere complexes with the concave side 5.0 Å from the layer midplane, with a second 'shared' position at 8.9 Å, approximately half the distance to the next layer (Fig. 3d-f). On the exterior side, we observe outer-sphere complexes at 6.4 Å that were lower in relative abundance compared to the interlayer. High contrast planes at 11.1 Å and 14.4 Å qualitatively agree with atomistic simulations showing an increase and subsequent periodic variation in the chlorine concentration[17,38], but stark differences in the sodium distributions on different sides of curved layers have not been presaged from prior experiment or theory.

In layer 2, sodium ions occupy the same inner-sphere, outer-sphere, and shared complexation sites on both sides, but the relative abundance and the lateral ordering in the interlayer also differs on convex and concave sides (Fig. 3g-i). We interpret this interfacial ion association that is largely conserved along a layer but differs between neighboring layers as waves of charge fluctuation[39] that are discretized to inner-sphere, outer-sphere and shared complexation configurations and are captured in images after being frozen. One key result of the 'complexation waves' we observe here is that each layer is polarized, and the complexed ions establish interfacial dipoles that interact with one another[36]. These complexation dipoles are generally not aligned normal to the plane of the layer, and therefore exert lateral forces. Acting together, the collective action of the complexation dipoles drives the bending of the layer, and conversely, the mechanical bending of a layer changes the complexation dipole interaction magnitude and direction. Thus, a new type of long-range electrostatic dispersion force emerges at the mineral-solution interface from the action of complexation waves.

Complexation waves alter the electrostatic potential of the curved mineral interface, which generates forces that extend tens of nanometers into solution via interactions with the electrolyte. This likely explains previous observations of long-range interactions between clay layers[40]. Furthermore, the frozen snapshots shown here reveal a broad distribution of layer and complexation configurations, suggesting that they may readily interconvert at ambient temperature, which we investigate *in situ* using X-ray photon correlation spectroscopy (XPCS).

Structural dynamics of Li-Mt and Na-Mt in 1 M electrolyte were investigated using the two-time correlation function of coherent scattering intensity from XPCS (Fig. 4a, b). Both systems exhibit correlations that ebb and flow, characteristic of non-equilibrium behavior, but their structures fluctuate over different timescales. At a scattering vector $q = 0.0025$ Å$^{-1}$, corresponding to the approximately 250 nm average diameter of an Mt layer, Li-Mt has arrested dynamics characteristic of a gel (Fig. 4a) in which the layers move very little over the course of 100 seconds.

The case of Na-Mt is qualitatively similar on much smaller timescales, but abrupt, avalanche transitions between correlated and uncorrelated states occur every few seconds. Transitions occur on the hundreds of millisecond timescales in 1 M potassium chloride solutions of potassium montmorillonite (K-Mt, Supporting Information). Based on the imaging above and the results of previous work[11], we interpret these avalanche transitions between gelled states as corresponding to the formation and exfoliation of stacks, which is inherently coupled to the ability of layers to bend and ions to move. Layer diffusivities are many orders of magnitude smaller than those of ions or water (Supporting Information) and therefore the abrupt transitions arise when the electrolyte adopts a configuration that is unstable but requires the mesoscale bending and restacking of layers to resolve. The surprising observation that Li-Mt has significantly slower dynamics than Na-Mt, and Na-Mt far slower than K-Mt, is attributable to the hydration free energy of cation, which slows the rate of interconversion between complexation configurations at the mineral interface and thus the propagation of complexation waves.

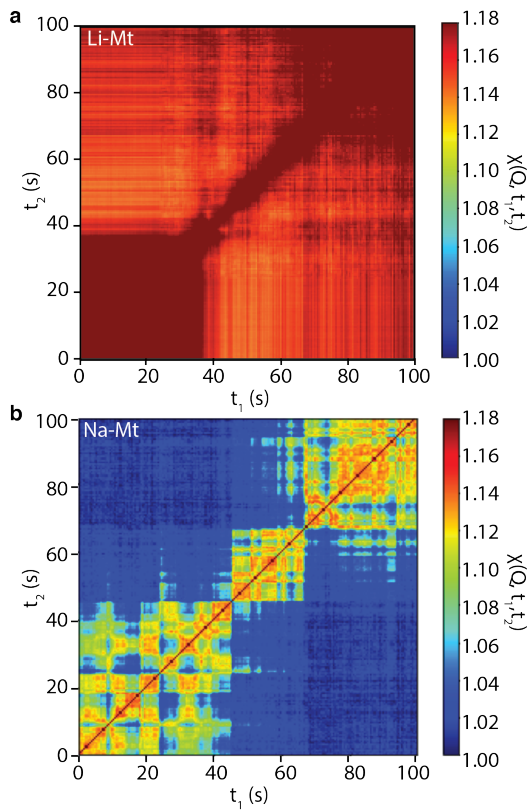

**Fig 4 | Two-time correlation plots of hydrated Mt.** (a) Li-Mt in 1 M lithium chloride. (b) Na-Mt in 1 M sodium chloride.

Together, these findings demonstrate that avalanche transitions are a general phenomenon in clay systems that arise from the exchange of elastic, electrostatic and hydration energy as ions partition from the bulk electrolyte, complex with the mineral layer and induce it to bend. A new interaction force emerges through the delocalizing effect of curvature, which spans length scales ranging from hundreds of nanometers to ångstroms, and strongly effects the temporal response of the system. CryoET and cryoEM do not yet resolve interfacial complexation configurations in three-dimensions, but nevertheless open a new window into the structure of hydrated interfaces. This direct observation is essential for the quantitative interpretation of aqueous interfacial phenomena that underlie myriad geochemical systems.

Supplementary Information is linked to the online version of the paper at [add link]

### Acknowledgements

This research was supported by the U.S. Department of Energy, Office of Science, Office of Basic Energy Sciences, Chemical Sciences, Geosciences, and Biosciences Division, through its Geoscience program at LBNL under Contract DE-AC02-05CH11231. Work at the Molecular Foundry was supported by the Office of Science, Office of Basic Energy Sciences, US. Department of Energy Contract DE-AC02-05CH11231. C.O. Acknowledges support from the U.S. Department of Energy Early Career Research Program. We thank Laura Lammers, Christophe Tournassat and Chenhui Zhu for helpful discussions. We thank Dan Toso for technical assistance with the FEI Titan Krios.

### Author Contributions

MLW conceived the study design, performed cryoEM and cryoET experiments, analyzed data, and developed alignment and segmentation algorithms. DR and LW developed alignment and reconstruction algorithms and DR analyzed the data cryoET. CO developed alignment, segmentation, and reconstruction algorithms and analyzed cryoEM and cryoET data. YZ performed XPCS experiments and analyzed the data. BG and JFB conceived the study design. All authors contributed to writing the manuscript.

### Author Information

Reprints and permissions information is available at [add link]
The authors declare no competing interests.
Correspondence and requests for materials should be addressed to mwhittaker@lbl.gov

### Methods

**Materials.** Wyoming montmorillonite (SWy-3), obtained from the Source Clays Repository of The Clay Minerals Society (http://www.clays.org/sourceclays_data.html), was used throughout this study. Aqueous solutions of lithium chloride and sodium chloride were prepared from reagent-grade salts.

**CryoET/cryoEM.** Suspensions of Li-Mt and Na-Mt with mineral concentrations of 5 mg/mL were deposited as 2-3 µL aliquots onto 300-mesh lacy carbon Cu grids (Electron Microscopy Sciences) which had been glow-discharged in air plasma for 15 seconds. Excess solution was removed by automatic blotting (1 blot for 10 s, blot force 10 at 95% relative humidity) before plunge-freezing in liquid ethane using an automated vitrification system (FEI Vitrobot). Imaging was performed with a Titan Krios TEM operated at 300 kV, equipped with a BIO Quantum energy filter. Images were recorded on a Gatan K3 direct electron detecting camera with a pixel size of 0.91 Å/pixel in superresolution mode for cryoET and 0.75 Å/pixel for cryoEM. Imaging was performed under cryogenic conditions using a low electron dose of 121 e$^-$/Å$^2$ for cryoET images and 1100 e$^-$/Å$^2$ for cryoET. Acquisition was automated with SerialEM software.

**CryoET reconstruction.** Dose-fractionated movies acquired at each tilt and defocus were gain corrected and aligned in RELION3.0 and summed to form intensity images. Intensity images were normalized, Fourier downsampled to 3.64 Å/pixel, pre-aligned using an in-house code written in MATLAB, and the alignment was refined using IMOD. Tomographic reconstruction was performed iteratively using the method developed by Ren et al.[35] to model HRTEM contrast from multiply scattering samples. We used custom implemented Python library, which supports GPU computation (https://github.com/yhren1993/PhaseContrastTomographySolver) to perform reconstructions on the Lawrence Berkeley National Lab High Performance Computing Clusters in a distributed fashion. The final reconstruction had an isotropic voxel size of 3.64 Å.

**CryoET segmentation.** The clay sheets in the reconstructed cryo-EM absorption volumes were segmented using custom codes written in Matlab. The goal was to reduce the set of voxels to a set of "sheets" which were defined as a cloud of points representing a 2D sheet embedded in 3D space, each with an associated normal vector representing the sheet surface normal. We used Fast Fourier Transforms to efficiently compute the correlation between an isotropic orientation kernel and each voxel in the reconstruction volume, then applied a two-level threshold (1) a global threshold by using a minimum value for the correlation signal, and (2) voxels with correlation signals greater than at least 18 neighboring voxels (out of a possible 26 neighbors). Mean surface curvature 2H was estimated by fitting the 2D parabolic local surface to the expression

$$2H = \frac{(1+S_x^2)S_{yy} - 2S_xS_yS_{xy} + (1+S_y^2)S_{xx}}{(1+S_x^2+S_y^2)^{3/2}}$$

Where $S_x$ and $S_y$ are the first order spatial derivatives of the surface, and $S_{xx}$, $S_{yy}$, and $S_{xy}$ are the second order derivatives.

**X-ray scattering.** X-ray scattering was performed at beamline 5ID-D of the Advanced Photon Source at Argonne National Laboratory in order to obtain high photon fluxes necessary for time-resolved experiments. Small-, medium-, and wide-angle X-ray scattering (SAXS/MAXS/WAXS) was collected simultaneously on three Rayonix charge-coupled device (CCD) detectors with sample−detector distances of 8505.0, 1012.1, and 199.5 mm, respectively. The wavelength of radiation was set to 1.2398 Å (10 keV), resulting in a continuous range of scattering vector, $q = 0.017-4.2$ Å$^{-1}$.

**XPCS.** XPCS experiments were conducted at the Coherent Hard X-ray (CHX) beamline 11-ID at the National Synchrotron Light Source II (NSLS-II), Brookhaven National Laboratory. The X-ray energy was 9.65 keV (λ= 1.285 Å) with energy resolution ΔE/E ≈10$^{-4}$ from a Si111 double crystal monochromator. A partially coherent X-ray beam with a flux at the sample of ~10$^{11}$ photons/sec and a focused beam size of 10 × 10 μm$^2$ was achieved by focusing with a set of Be Compound Refractive Lenses and a set of Si kinoform lenses in front of the sample. The sample was loaded in a wax-sealed glass capillary mounted on the sample stage. The coherent scattering pattern was recorded in transmission small angle scattering geometry by using a photon-counting pixelated area detector (Eiger X 4M Dectris Inc.) located 16.03 meters away from the sample with a 75 μm × 75 μm pixel size. The X-ray radiation dose on the sample was controlled by a millisecond shutter and filters of different thickness of silicon wafers. The data acquisition strategy was optimized to ensure that the measured dynamics and structure are dose independent. The XPCS data analysis were conducted by using software developed at CHX, NSLS-II. A $q$ range of $Q = 0.0015 - 0.9$ Å$^{-1}$, corresponding to length scales of 0.7-420 nm, captures the lateral dimensions and interlayer spacings of all layers. The two-time correlation function, $\chi(Q, t_1, t_2)$, equals unity if there is no correlation between X-ray scattering intensities at a scattering vector $Q$ for an initial time $t_1$ and second time $t_2$, and approaches 1.2 for intensities that are unchanged between the two time points.

# Supporting Information for "Dynamic clay microstructures evolve via emergent ion complexation waves"

Michael L. Whittaker,[1,2] David Ren,[3] Colin Ophus,[4] Yugang Zhang,[5] Benjamin Gilbert,[1,2] Laura Waller[3], Jillian F. Banfield[1,2]

[1] Energy Geosciences Division, Lawrence Berkeley National Laboratory, Berkeley, California, USA 94720.
[2] Department of Earth and Planetary Science, University of California, Berkeley, California, USA 94720.
[3] Department of Electrical Engineering and Computer Sciences, University of California, Berkeley, California, USA 94720.
[4] National Center for Electron Microscopy, Molecular Foundry, Lawrence Berkeley National Laboratory, Berkeley, California, USA 94720.
[5] Brookhaven National Laboratory, Upton, New York, USA 11973.

# 1. Suspension microstructures
## 1.1. CET

The clay sheets in the reconstructed cryo-ET absorption volumes were segmented using custom codes written in Matlab. The below variables were defined for a bin 2 reconstruction size. The goal was to reduce the set of voxels to a set of "sheets" which were defined as a cloud of points representing a 2D sheet embedded in 3D space, each with an associated normal vector representing the sheet surface normal. First, we generated a list of unit vectors $\boldsymbol{n_i}$ with roughly even angular spacing over a hemispherical surface, representing the possible sheet orientations. Next, for each orientation we generated a 3D kernel $\boldsymbol{k_i}$ defined by the function

$$\boldsymbol{k_i} = \frac{1}{2} - \frac{1}{2}\mathrm{erf}\left(\frac{t/2 - |\boldsymbol{r} \cdot \boldsymbol{n_i}|}{w}\right) \tag{1}$$

where $\boldsymbol{r}$ is the real space coordinates centered on the origin, $t = 3$ is the estimated sheet thickness, and $w = 1$ is the estimated sheet interfacial width. This kernel was normalized by applying a 3D Gaussian envelope function and subtracting the mean, i.e. the formula

$$\boldsymbol{k_i}^{\mathrm{norm}} = \left[\boldsymbol{k_i} - \langle \boldsymbol{k_i} \exp\left(-\frac{|\boldsymbol{r}|^2}{2\sigma^2}\right)\rangle\right] \exp\left(-\frac{|\boldsymbol{r}|^2}{2\sigma^2}\right) \tag{2}$$

where $\sigma = 8$ is the Gaussian envelope standard deviation, and $\langle \ \rangle$ represents the mean or expectation value. For each potential orientation, we use Fast Fourier Transforms to efficiently compute the correlation of the kernel with the reconstructed volume. We then take the maximum correlation value in each voxel over all orientations, while also storing the best-match orientation. We then classified the sheet voxels by applying two thresholds: (1) a global threshold by using a minimum value for the correlation signal, and (2) voxels with correlation signals greater than at least 18 neighboring voxels (out of a possible 26 neighbors).

Next, we computed the nearest neighbor network for the set of all sheet voxels. This network was used to segment the set into separate sheets using two matching rules: (1) locally connected voxels, and (2) those with orientations within 30° of each other. At this stage we also discarded sheets which consisted of less than 1000 voxels, since these were either false positives or sheets too small to make accurate measurements of the surface topology.

Finally, we refined the sheet voxel coordinates and surface normal using an iterative procedure. The first step consisted of moving a voxel to the center position of a parabolic surface fitted to all of the neighboring sheet voxels within 55 voxels. Next, the sheet orientation was updated to be the normal of this parabolic surface at the fitted voxel position. Finally, the voxel was moved perpendicular to the surface (along the orientation normal vector) to the region of lowest absorption signal in the original reconstructed volume. This process was used to remove the "pixelization" of the originally detected coordinate positions. We repeated these steps until all points had converged.

The final analysis steps consisted of measurements performed on the segmented surfaces. Line traces were defined using the surface normal vectors at each point. Mean surface curvature $2H$ was estimated by fitting the 2D parabolic local surface to the expression

$$2H = \frac{(1+S_x^2)S_{yy} - 2S_xS_yS_{xy} + (1+S_y^2)S_{xx}}{(1+S_x^2+S_y^2)^{3/2}} \quad (3)$$

Where $S_x$ and $S_y$ are the first order spatial derivatives of the surface, and $S_{xx}$, $S_{yy}$, and $S_{xy}$ are the second order derivatives.

Subvolumes from reconstructed tomograms in Fig. 1, showing average interlayer spacings of curved osmotic hydrate stacks that contribute to the cryoET structure factor are given in Fig. S1.

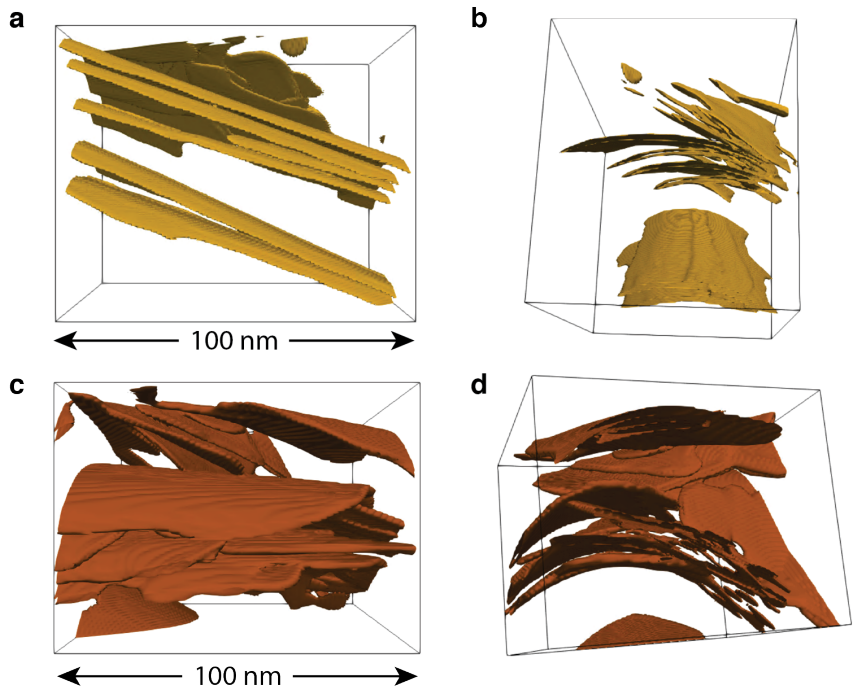

Figure S1. Subvolumes from cryoET reconstructions in Fig. 1. (a, b) Subvolume from Fig. 1a, showing stacked layers with coaligned curvature axes and variable interlayer spacing. (c, d) Subvolume from Fig. 1d, showing stacked layers with smaller average interlayer spacing and smaller radius of curvature.

## 1.2. X-ray scattering

Clay mineral microstructures are often inferred from their average interlayer distances, $<D>$, which in the simplest case of a nematic phase with planar and parallel layers, can be determined from

$$\langle D \rangle = t\phi^{-1} \tag{4}$$

Equation (4) represents the maximum average interlayer spacing for a given mineral volume fraction, $\phi$, and is valid above $\phi = 2\%$ for Li-Mt (Fig. S2)

The value of $<D>$ is typically found from the structure factor, $Iq^2$, in suspension determined by small-angle X-ray scattering. Structure factors exhibit a transition from nematic to isotropic structures at elevated electrolyte concentrations (Figure S3a) and $<D>$ increases more slowly with increasing mineral volume fraction than Equation (4) predicts (Fig. S2a). A heuristic relationship for $<D>$ has been shown to apply[1] over a range of clay fractions

$$\langle D \rangle = \frac{l}{\left(\phi \frac{2l}{3t}\right)^{1/3} + \left(\phi \frac{2l}{3t}\right)} \tag{5}$$

The general applicability of Equation (5) indicates that the harmonic mean of nematic and isotropic structures satisfactorily describes the average layered mineral structure at large length scales. As we show here, the structures are not isotropic at length scales below one micrometer. However, Equation (5) implicitly captures the combined effects of bending (through $\phi^{1/3}$ dependence), stacking (through $\phi^1$ dependence) and the dynamic interplay between the two.

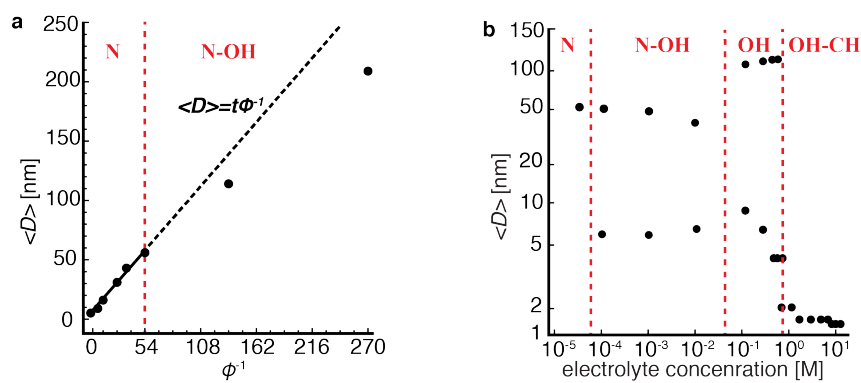

Figure S2. Average interlayer spacing from X-ray scattering structure factors (a) as a function of decreasing mineral volume fraction and (b) as a function of increasing electrolyte concentration. Nematic (N) structures are prevalent at low electrolyte concentration and high volume fraction, while osmotic hydrates (OH) of curved layers appear as the electrolyte concentration increases or mineral volume fraction decreases. Crystalline hydrates (CH) appear at high electrolyte concentrations.

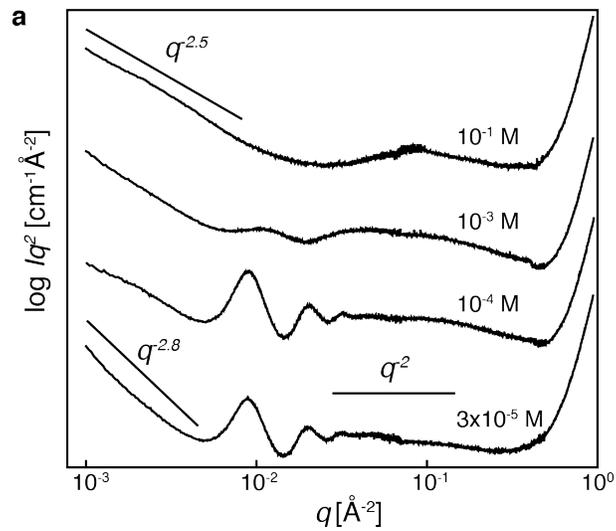

Figure S3. X-ray scattering from Li-Mt at various electrolyte concentrations. In the absence of background electrolyte ($3\times 10^{-5}$ M) the structure factor.

## 2. Suspension dynamics

### 2.1. XPCS

Suspension dynamics determined by XPCS varied widely between Li-Mt, Na-Mt and K-Mt. Correlation between coherent scattering intensity at $q = 0.0025$ Å$^{-1}$, corresponding to the approximately 250 nm average diameter of an Mt layer, is shown in Fig. S4. At equivalent mineral (2%) and alkali chloride electrolyte (1 M) concentrations, decreasing cation hydration energy[2] was directly correlated with faster dynamics, which manifest as transitions between correlated (1.18) and uncorrelated (1.00) states.

Equilibrium fluctuations between two-dimensional speckle patterns from coherent X-ray scattering were quantified using the (one-time) intensity autocorrelation function, $g_2(q, t)$, defined at each pixel as

$$g_2(q,t) = \frac{\langle I(q,t)I(q,t+\tau)\rangle_T}{\langle I(q,\tau)\rangle_T^2} \quad (6)$$

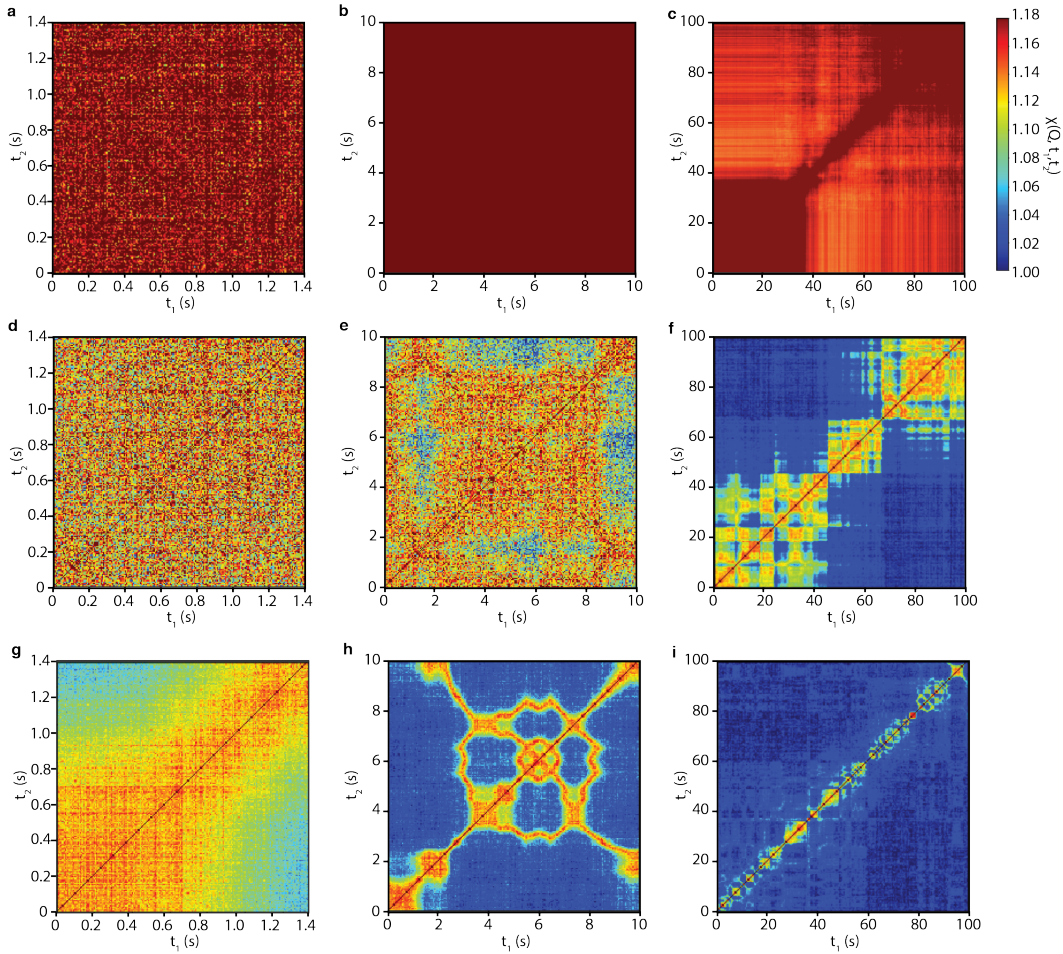

Figure S4. XPCS two-time correlation plots. (a-c) Li-Mt, (d-f) Na-Mt, (g-i) K-Mt.

where $I(q, t+\tau)$ is the intensity at scattering vector $q$ and time $t+\tau$ for a given timepoint $t$ and $T$ is the duration of acquisition. Values of $g_2(q, t)$ are averaged over a range of $q$ values corresponding to a length scale of interest. For the beamline optics used in this study, 1.18 is the maximum value of $g_2(q, t)$ and a minimum value of 1.0 corresponds to completely uncorrelated structures.

The two-time correlation function, $\chi(q, t_1, t_2)$, is analogously used to define the intensity autocorrelation when the relaxation time constant changes over time. In this non-equilibrium case

$$\chi(q, t_1, t_2) = \frac{I(q,t_1)I(q,t_2)}{\langle I(q,t)\rangle^2} \quad (7)$$

The change in $g_2(q, t)$ with time determined from both one- and two-time intensity autocorrelation functions was fit with a stretched exponential of the form

$$g_2(q, \tau) = 1 + \beta(q)exp(-2D(q)q^2\tau) \quad (8)$$

where $\beta(q)$ is the $q$-dependent contrast magnitude and $D(q)$ is the $q$-dependent diffusion coefficient. Diffusion coefficients, $D_o$, were calculated from Equation (8) and found to be highly dependent on the counterion identity. Diffusion coefficients of K-Mt were nearly three orders of magnitude larger than Na-Mt and six orders of magnitude larger than Li-Mt.

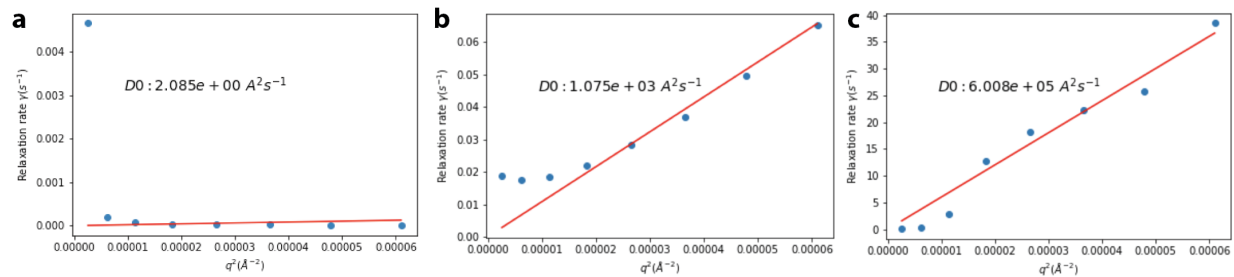

Figure S5. Diffusion coefficients calculated from the size-dependent relaxation time. (a) Li-Mt, (b) Na-Mt, (c) K-Mt.